\documentclass{elsart3-1}

\usepackage{amssymb}

\usepackage[english,francais]{babel}

\newtheorem{e-proposition}[theorem]{Proposition}

\newtheorem{e-definition}[theorem]{Definition\rm}

\newtheorem{theoreme}{Th\'eor\`eme}[section]

\newtheorem{proposition}[theoreme]{Proposition}

\newtheorem{remarque}{\it Remarque}

\renewcommand{\theequation}{\arabic{equation}}
\def\og{\leavevmode\raise.3ex\hbox{$\scriptscriptstyle\langle\!\langle$~}}
\def\fg{\leavevmode\raise.3ex\hbox{~$\!\scriptscriptstyle\,\rangle\!\rangle$}}

\begin{document}
% Vous pouvez mettre dans la prochain ligne la rubrique choisie
% (si vous la connaissez) - meme deux, format : Rubrique1/Rubrique2
\centerline{Th\'{e}orie des signaux / Automatique th\'{e}orique}
\begin{frontmatter}

% Titre, auteurs et adresses

% utiliser la commande \thanksref dans \title, \author ou \address
%     pour les notes en bas de page ;

% utiliser la commande \ead pour l'adresse e-mail de chaque auteur
%    (apr\"{E}s la commande \auteur) ;

% \title{Title\thanksref{label1}}
% \thanks[label1]{}
% \author{Name\thanksref{label2}}
% \ead{email address}
%
% \thanks[label2]{}
% \address{Address\thanksref{label3}}
% \thanks[label3]{}
\selectlanguage{francais}
\title{Analyse non standard du bruit}

% utiliser les \`{E}tiquettes pour indiquer l'adresse de chaque auteur,
%     s'il y a plusieurs adresses

% \author[label1,label2]{}
% \address[label1]{}
% \address[label2]{}

\author[authorlabel1]{Michel FLIESS},
\ead{Michel.Fliess@polytechnique.edu}
%\author[authorlabel2]{Auteur Nom2}
%\ead{auteur.nom2@email.adresse}

\address[authorlabel1]{Projet ALIEN, INRIA Futurs \\ \& \'Equipe MAX, LIX (CNRS, UMR 7161),
\'Ecole polytechnique, 91128 Palaiseau, France }
%\address[authorlabel2]{Adresse2}
% etc, etc

% Vous pouvez mettre a la prochaine ligne les dates
% (de reception et d'acceptation), et le nom du presentateur de votre Note

\medskip
\selectlanguage{francais}
\begin{center}
{\small Re\c{c}u le *****~; accept\'e apr\`es r\'evision le +++++\\
Pr\'esent\'e par Yves Meyer}
\end{center}

\begin{abstract}
% resume en francais, et apres l'abstract en anglais, qui
%    commence avec le titre en gras.
\selectlanguage{francais}
% Texte du r\`{E}sum\`{E} en fran\'{A}ais
La formalisation non standard, due \`{a} P. Cartier et Y. Perrin, des
oscillations rapides fournit un cadre math\'{e}matique ad\'{e}quat pour de
nouvelles techniques d'estimation non asymptotiques, ne n\'{e}cessitant
pas l'analyse statistique habituelle des bruits entachant tout
capteur. On en tire diverses cons\'{e}quences sur les bruits
multiplicatifs, la largeur des fen\^etres d'estimations
param\'{e}triques et les erreurs en rafales. {\it Pour citer cet
article: M. Fliess, C. R. Acad. Sci. Paris, Ser. I 342 (2006).}
\vskip 0.5\baselineskip

\selectlanguage{english} \noindent{\bf Abstract} \vskip
0.5\baselineskip \noindent {\bf Noises: a nonstandard analysis.}
Thanks to the nonstandard formalization of fast oscillating
functions, due to P. Cartier and Y. Perrin, an appropriate
mathematical framework is derived for new non-asymptotic estimation
techniques, which do not necessitate any statistical analysis of the
noises corrupting any sensor. Various applications are deduced for
multiplicative noises, for the length of the parametric estimation
windows, and for burst errors. {\it To cite this article: M. Fliess,
C. R. Acad. Sci. Paris, Ser. I 342 (2006).}
\end{abstract}
\end{frontmatter}

% Maintenant la version abr\`{E}g\`{E}e en anglais, si pr\`{E}sente
\selectlanguage{english}
\section*{Abridged English version}
% Texte de la version abr\`{E}g\`{E}e en anglais
\section{Introduction}
Recent works (see, {\it e.g.},
\cite{gretsi2,cras,gretsi,cifa,zeitz,fmmsr,esaim,cras1}) in signal
processing and automatic control yield quite efficient estimation
techniques, which are non-asymptotic and do not necessitate any
statistical treatment of the noises corrupting any sensor. An
appropriate mathematical framework is provided by the nonstandard
formalization, which is due to Cartier and Perrin \cite{cartier}, of
fast oscillating, or fluctuating, functions. Various applications
are derived for multiplicative noises, for the lengths of the
estimation windows, and for burst errors.

\section{Noises}
Let $^*\mathbb{N}$, $^*\mathbb{R}$ be the usual nonstandard
extensions of $\mathbb{N}$, $\mathbb{R}$. Replace $[0, 1]^{1 + D}
\subset {\mathbb{R}}^{1 + D}$ by the hyperfinite set ${\mathrm{I}} =
{\mathrm{I}}_{0} \times {\mathrm{I}}_1 \times \dots \times
{\mathrm{I}}_D$, ${\mathrm{I}}_\iota = \{0, \frac{1}{N_\iota},
\dots, \frac{N_\iota - 1}{N_\iota}, 1 \}$, where $N_\iota \in
{^*\mathbb{N}}$, $\iota = 0, 1, \dots, D$, $D \in \mathbb{N}$, is
unlimited. The function $\lambda_\iota: {\mathrm{I}}_\iota
\backslash \{1\} \rightarrow {^*\mathbb{R}}$,
$\frac{k_\iota}{N_\iota} \mapsto \frac{1}{N_\iota}$, $0 \leq k_\iota
\leq N_\iota - 1$, is the {\em Lebesgue measure} of
${\mathrm{I}}_\iota$ \cite{cartier}. The Lebesgue measure $\lambda =
\lambda_0 \times \lambda_1 \times \dots \times \lambda_D$ of
${\mathrm{I}}$ is clear. See \cite{cartier} for the notions of {\em
quadrable} sets and of {\em Lebesgue integrability}. A
$S$-integrable function $f: {\mathrm{I}} \rightarrow {^*\mathbb{R}}$
is {\em fast oscillating} \cite{cartier}, or {\em fast fluctuating},
if, and only if, any integral $\int_A f d\lambda$, where $A
\subseteq {\mathrm{I}}$ is quadrable, is infinitesimal. Let $m:
{\mathrm{I}} \rightarrow {^*\mathbb{R}}$ be a Lebesgue integrable
function. A {\em noise}, of {\em mean} $m$, is a $S$-integrable
function $\mathfrak{n}: {\mathrm{I}} \rightarrow {^*\mathbb{R}}$
such that, $\forall ~ \iota = 0, 1, \dots, D$, $\forall ~ \xi_\kappa
\in {\mathrm{I}}_\kappa$, $\kappa \neq \iota$, the projection
${\mathfrak{n}}(\xi_0, \xi_1, \dots, \xi_D) - m (\xi_0, \xi_1,
\dots, \xi_D): {\mathrm{I}}_\iota \rightarrow {^*\mathbb{R}}$ is
fast oscillating. If it is possible to take $m \equiv 0$,
$\mathfrak{n}$ is said to be {\em zero-mean}.

\section{Applications}
\subsection{Multiplicative and additive noises}
Let $x: {\mathrm{I}} \rightarrow {^*\mathbb{R}}$ be a function of
class $S^1$ \cite{diener}, which we call a {\em signal}. Let the
sensor $y = {\mathfrak{n}}_1 x + {\mathfrak{n}}_2$ be corrupted by a
multiplicative noise ${\mathfrak{n}}_1$, of mean $1$, and an
additive noise ${\mathfrak{n}}_2$, of arbitrary mean.
\begin{proposition}
$y = x  + {\mathfrak{n}}$ where ${\mathfrak{n}} = {\mathfrak{n}}_2 +
({\mathfrak{n}}_1 - 1)x$ is a noise of the same mean as
${\mathfrak{n}}_2$.
\end{proposition}

\subsection{Parametric estimations}
Set $D = 0$. Assume that the the shadow of the signal $x:
{\mathrm{I}}_0 \rightarrow {\mathbb{R}}$ is analytic. Assume also
that the constant parameter $\theta \in \mathbb{R}$ is {\em linearly
identifiable} \cite{cras,fmmsr}. It leads to the estimator
(\ref{estim}) where $[ \theta ]_e (t)$ is the estimate of $\theta$
at $t$;
%\begin{itemize}
$[0, t]$ is the {\em estimation window} of {\em length} $t$; the
shadow $\bar{\delta} (t)$ of $\delta (t)$ is an analytic function
$[0, 1] \rightarrow \mathbb{R}$, called {\em divisor}, such that
$\bar{\delta} (0) = 0$; $\mathfrak{n}$ is an additive noise.
%\end{itemize}

%\subsubsection{Windows lengths}
\begin{proposition}\label{zero}
Assume that $\mathfrak{n}$ is zero-mean. If the length of the
estimation window does not belong to the monads of the divisor's
zeros, the estimate $[ \theta ]_e (t)$ belongs to the monad of
$\theta$. The length is then appreciable. If the length is
infinitesimal, $[ \theta ]_e (t)$ does not belong necessarily to the
monad of $\theta$.
\end{proposition}

%\subsubsection{Demodulation and burst errors}
The parameter, often called {\em symbol}, which has to be {\em
demodulated} (see, {\it e.g.}, \cite{code,proakis}) is generally
associated to a signal satisfying a linear differential equation
with time-polynomial coefficients. {\em Burst errors} may be
understood in our setting in the following way: contrarily to
Proposition \ref{zero} the noise is not zero-mean. Since the
transmission duration of any signal is short, we may assume that the
shadow of the unknown mean $m(t)$ is a polynomial $p(t)$ of a given
limited degree. Consider $p$ as a structured perturbation which may
be annihilated \cite{cras,fmmsr} by some given limited power of
$\frac{d}{dt}$.
\begin{proposition}
Assume that the shadow of the signal is not annihilated by a limited
power of $\frac{d}{dt}$. There exists then an estimate $[ \theta ]_e
(t)$ of the symbol which belongs to the monad of $\theta$, for any
$t$ which does not belong to the monad of a divisor's zero.
\end{proposition}

\setcounter{section}{0} \selectlanguage{francais}
% texte principale
\section{Introduction}
\label{intro} Des travaux r\'{e}cents (voir, par exemple,
\cite{gretsi2,cras,gretsi,cifa,zeitz,fmmsr,esaim,cras1}) conduisent
\`{a} des techniques d'estimation non asymptotiques, efficaces en signal
et en automatique, sans aucun recours aux traitements statistiques
habituels des bruits entachant tout capteur. Rappelons bri\`{e}vement de
quoi il retourne pour ces m\'{e}thodes, illustr\'{e}es par maints exemples,
mal ma\^{\i}tris\'{e}s par ailleurs. On distingue
\cite{cras,fmmsr,esaim,cras1} deux types de perturbations, celles
dites {\em structur\'{e}es}, annihil\'{e}es avec des op\'{e}rateurs
diff\'{e}rentiels lin\'{e}aires, et celles dites {\em non structur\'{e}es},
consid\'{e}r\'{e}es comme des oscillations, ou fluctuations, rapides,
att\'{e}nu\'{e}es par des filtres passe-bas, comme les int\'{e}grales it\'{e}r\'{e}es.

La formalisation non standard \cite{robinson} des oscillations
rapides, due \`a Cartier et Perrin \cite{cartier}, g\'{e}n\'{e}ralisant des
travaux ant\'{e}rieurs de Harthong \cite{harthong} et Reder
\cite{reder}, fournit un cadre math\'{e}matique, permettant d'introduire
les bruits et leurs moyennes dans un cadre enti\`{e}rement d\'{e}terministe.
Plusieurs justifications issues de la pratique des ing\'{e}nieurs sont
propos\'{e}es. On en d\'{e}duit divers r\'{e}sultats sur les bruits
multiplicatifs, la largeur des fen\^etres d'estimations, que
confortent d\'{e}j\`{a} de nombreuses simulations num\'{e}riques (voir
\cite{gretsi2,cras,gretsi,cifa,zeitz,fmmsr,esaim,cras1} et leurs
bibliographies) et quelques exp\'{e}riences de laboratoire, et sur les
erreurs en rafales, {\em burst errors} en am\'{e}ricain, pour lesquelles
ce travail reste \`{a} faire.

\section{Bruits}
Nous avons r\'{e}dig\'{e} selon le formalisme {\it ZFC} de Robinson
\cite{robinson}, tout en empruntant beaucoup au vocabulaire de
\cite{cartier} et \cite{diener}, qui se situent dans l'{\it IST} de
Nelson \cite{bams}.
%Le lecteur \'{e}tranger \`{a} cette analyse trouvera en
%\cite{cnrs} un excellent panorama historique et \'{e}pist\'{e}mologique.

\subsection{D\'{e}finition non standard}\label{bsn}
D\'{e}signons par $^*\mathbb{N}$, $^*\mathbb{R}$ les extensions non
standard de $\mathbb{N}$, $\mathbb{R}$. Rempla\c{c}ons $[0, 1]^{1 + D}
\subset {\mathbb{R}}^{1 + D}$ par l'ensemble hyperfini ${\mathrm{I}}
= {\mathrm{I}}_{0} \times {\mathrm{I}}_1 \times \dots \times
{\mathrm{I}}_D$, ${\mathrm{I}}_\iota = \{0, \frac{1}{N_\iota},
\dots, \frac{N_\iota - 1}{N_\iota}, 1 \}$, o\`u $N_\iota \in
{^*\mathbb{N}}$, $\iota = 0, 1, \dots, D$, $D \in \mathbb{N}$, est
illimit\'{e}. La fonction $\lambda_\iota: {\mathrm{I}}_\iota \backslash
\{1\} \rightarrow {^*\mathbb{R}}$, $\frac{k_\iota}{N_\iota} \mapsto
\frac{1}{N_\iota}$, $0 \leq k_\iota \leq N_\iota - 1$, est appel\'{e}e
\cite{cartier} {\em mesure de Lebesgue} de ${\mathrm{I}}_\iota$. On
d\'{e}finit alors, de fa\c{c}on \'{e}vidente, la mesure de Lebesgue $\lambda =
\lambda_0 \times \lambda_1 \times \dots \times \lambda_D$ de
${\mathrm{I}}$. Renvoyons \`a \cite{cartier} pour les notions
d'ensembles {\em quadrables} et de {\em Lebesgue-int\'{e}grabilit\'{e}}. Une
fonction $S$-int\'{e}grable $f: {\mathrm{I}} \rightarrow {^*\mathbb{R}}$
est dite \`{a} {\em oscillations}, ou {\em fluctuations}, {\em rapides}
\cite{cartier} si, et seulement si, toute int\'{e}grale $\int_A f
d\lambda$ est infinit\'{e}simale, o\`{u} $A \subseteq {\mathrm{I}}$ est
quadrable.

La d\'{e}finition suivante paraphrase le th\'{e}or\`{e}me 9.3.6 de
\cite{cartier} sur la d\'{e}composition d'une fonction $S$-int\'{e}grable en
somme d'une fonction Lebesgue-int\'{e}grable et d'une fonction
$S$-int\'{e}grable \`{a} oscillations rapides. Soit $m: {\mathrm{I}}
\rightarrow {^*\mathbb{R}}$ une fonction Lebesgue-int\'{e}grable. Un
{\em bruit}, de {\em moyenne} $m$, ou, en adaptant la terminologie
expressive de \cite{diener}, {\em cr\'{e}pitant} autour de $m$, est une
fonction $S$-int\'{e}grable $\mathfrak{n}: {\mathrm{I}} \rightarrow
{^*\mathbb{R}}$ telle que, $\forall ~ \iota = 0, 1, \dots, D$,
$\forall ~ \xi_\kappa \in {\mathrm{I}}_\kappa$, $\kappa \neq \iota$,
la projection ${\mathfrak{n}}(\xi_0, \xi_1, \dots, \xi_D) - m
(\xi_0, \xi_1, \dots, \xi_D): {\mathrm{I}}_\iota \rightarrow
{^*\mathbb{R}}$ est \`{a} oscillations rapides. La moyenne n'est,
\'{e}videmment, pas unique: $m^\prime$ est aussi une moyenne, si,
$\forall ~ \iota = 0, 1, \dots, D$, $\forall ~ \xi_\kappa \in
{\mathrm{I}}_\kappa$, $\kappa \neq \iota$, la projection $m^\prime
(\xi_0, \xi_1, \dots, \xi_D) - m (\xi_0, \xi_1, \dots, \xi_D):
{\mathrm{I}}_\iota \rightarrow {^*\mathbb{R}}$ est infinit\'{e}simale
{\em presque partout} (voir la proposition 9.3.12 de
\cite{cartier}). Si l'on peut prendre $m \equiv 0$, on dit que
$\mathfrak{n}$ est de {\em moyenne nulle}, ou {\em centr\'{e}}. Ce qui
suit est ais\'{e}:
\begin{proposition}\label{mult}
Soit $\phi: {\mathrm{I}} \rightarrow {^*\mathbb{R}}$ une fonction de
classe $S^1$ \cite{diener}. Alors, le produit $\phi \mathfrak{n}$
est encore un bruit, de moyenne $\phi m$. Si $\mathfrak{n}$ est
centr\'{e}, $\phi \mathfrak{n}$ l'est aussi.
\end{proposition}

\subsection{Justifications}
\subsubsection{Hautes fr\'{e}quences}
Soit $\underline{n} = \sum_{\tiny \mbox{\rm finie}} A \sin (\Omega t
+ \varphi)$, $A, \Omega, \varphi, t \in \mathbb{R}$. L'int\'{e}grale
$\int_{t_i}^{t_f} \underline{n}(\tau) d\tau$, $t_i, t_f \in
\mathbb{R}$, est \og petite \fg ~ avec de \og hautes \fg ~fr\'{e}quences
$\Omega$.
\begin{remarque}
Par contre, l'int\'{e}grale $\int_{t_i}^{t_f} (\underline{n}(\tau))^2
d\tau$ n'est pas \og petite \fg. L'{\em \'{e}cart-type} d'un bruit, au
sens du {\S} \ref{bsn}, qui est, avec sa moyenne, de carr\'{e} int\'{e}grable,
est donc, en g\'{e}n\'{e}ral, appr\'{e}ciable.
\end{remarque}

\subsubsection{Bruits blancs}
Consid\'{e}rons, avec bien des ouvrages pour ing\'{e}nieurs en traitement du
signal ({\it cf.} \cite{proakis}), le capteur $y$ d'un signal $x$
bruit\'{e} additivement
\begin{equation}\label{cont}
y(t) = x(t) + \bar{n} (t)
\end{equation}
$t \in [0, 1]$, o\`u $\bar{n} (t)$ est un bruit blanc, souvent
suppos\'{e} stationnaire, centr\'{e} et gaussien. Le lien avec le {\S}
\ref{bsn} d\'{e}coule de l'analogue \'{e}chantillonn\'{e}, usuel chez les
praticiens ({\it cf.} \cite{proakis}),
\begin{equation}\label{discr}
y (\alpha \Delta t) = x(\alpha \Delta t) + \bar{n}(\alpha \Delta t)
\end{equation}
$\alpha = 0, 1, \dots, \bar{N}$, $\bar{N} \in
{\mathbb{N}}$, o\`u
\begin{itemize}
\item $\Delta t = \frac{1}{\bar{N}}$ est le pas d'\'{e}chantillonnage,
limit\'{e} et appr\'{e}ciable,
\item les $\bar{n}(\alpha \Delta t)$ sont des variables al\'{e}atoires ind\'{e}pendantes,
centr\'{e}es, d'\'{e}carts-types normalis\'{e}s \`{a} $1$.
\end{itemize}
En effet, toute somme \begin{equation}\label{centlim} \frac{t_F -
t_I}{\bar{N}} \sum_{0 \leq t_I \leq \alpha \Delta t \leq t_F \leq 1}
{\bar{n}} (\alpha \Delta t) \end{equation} tend presque s\^urement
vers $0$ avec le pas d'\'{e}chantillonnage.

\begin{remarque}
Les travaux plus math\'{e}matiques, comme \cite{ibra}, substituent \`{a}
(\ref{cont}) l'\'{e}quation diff\'{e}rentielle stochastique $ dy = x dt +
dw$, o\`u $w$ est un processus de Wiener. La r\'{e}\'{e}criture $y(t) = y(0)
+ \int_{0}^{t} x(\tau)d\tau + w(t) - w(0)$ confirme, si besoin est,
que l'on ne peut se ramener \`{a} (\ref{cont}). Cette dichotomie sera
examin\'{e}e prochainement, de m\^eme que ses liens avec certaines
questions de {\em m\'{e}canique stochastique} ({\it cf.}
\cite{nelson-quant,not}), en nous inspirant des {\em moyennes
glissantes} de \cite{harthong,reder} et de la {\em d\'{e}rivation
quantique} de \cite{cresson}.
\end{remarque}

\section{Applications}
\subsection{Bruits additifs et multiplicatifs}
Soit $x: {\mathrm{I}} \rightarrow {^*\mathbb{R}}$ une fonction,
appel\'{e}e signal, suppos\'{e}e de classe $S^1$, ayant, par cons\'{e}quent, une
ombre $C^1$ \cite{diener}. Le capteur $y: {\mathrm{I}} \rightarrow
{^*\mathbb{R}}$ de $x$ est dit bruit\'{e} additivement et
multiplicativement si, et seulement si, $y = {\mathfrak{n}}_1 x  +
{\mathfrak{n}}_2$, o\`u les bruits multiplicatif ${\mathfrak{n}}_1$
et additif ${\mathfrak{n}}_2$ sont de moyennes respectives $1$ et
quelconque, suppos\'{e}e, souvent, nulle. Le r\'{e}sultat suivant, qui
repose sur le fait que $({\mathfrak{n}}_1 - 1) x$ est, d'apr\`{e}s la
proposition \ref{mult}, un bruit centr\'{e}, d\'{e}montre que l'on peut se
ramener au cas purement additif. On r\'{e}pond ainsi \`{a} des questions
courantes de la litt\'{e}rature appliqu\'{e}e ({\it cf.} \cite{isr,lim}) sur
la mani\`{e}re de traiter les bruits multiplicatifs.
\begin{proposition}
On peut \'{e}crire $y = x  + {\mathfrak{n}}$, o\`u ${\mathfrak{n}} =
{\mathfrak{n}}_2 + ({\mathfrak{n}}_1 - 1)x$ est un bruit de m\^eme
moyenne que ${\mathfrak{n}}_2$.
\end{proposition}

\subsection{Estimations param\'{e}triques}
Nos estimations pour les images et les vid\'{e}os \'{e}tant faites par
balayage unidimensionnel \cite{gretsi2}, il est loisible de poser $D
= 0$. Au vu des nombreux exemples de la litt\'{e}rature ({\it cf.}
\cite{proakis} et \cite{gretsi2,cras,gretsi,fmmsr}), on suppose que
le signal $x$ poss\`{e}de une ombre dans $C^\omega ([0, 1],
\mathbb{R})$. Soit $\theta \in \mathbb{R}$ un param\`{e}tre constant,
suppos\'{e} {\em lin\'{e}airement identifiable} \cite{cras,fmmsr}.
L'adaptation des calculs de \cite{cras,fmmsr} permet d'\'{e}crire un
estimateur sous la forme
\begin{equation}\label{estim}
\delta  (t) \left( [ \theta ]_e (t) - \theta \right) =
\sum_{\tiny{\mbox{\rm finie}}} c \int_{0}^{t} \dots
\int_{0}^{\tau_2} \int_{0}^{\tau_1} \tau_{1}^{\nu} \mathfrak{n}
(\tau_1)d\tau_1 d\tau_2 \dots d\tau_k
\end{equation}
o\`u
\begin{itemize}
\item $t \in {\mathrm{I}}_0$, $c \in \mathbb{R}$, $\nu \geq 0$, $k \geq
1$ sont limit\'{e}s,
\item $[0, t]$ est la {\em fen\^etre d'estimation}, de {\em largeur}
$t$,
\item $\delta (t)$ a pour ombre une fonction de $C^\omega ([0, 1],
\mathbb{R})$, appel\'{e}e {\em diviseur}, nulle en $0$,
\item $\mathfrak{n}$ est un bruit additif,
\item $[ \theta ]_e (t)$ est l'estim\'{e}e de $\theta$ en $t$.
\end{itemize}

\subsubsection{Largeur des fen\^etres}
Supposons $\mathfrak{n}$ centr\'{e}. On d\'{e}duit de (\ref{estim}) et de la
proposition \ref{mult} le r\'{e}sultat suivant qui corrobore le
caract\`{e}re non satistique et non asymptotique, mais non instantan\'{e}e,
de notre estimateur:
\begin{proposition}\label{fen}
Si la largeur de la fen\^etre d'estimation n'appartient pas au halo
d'un z\'{e}ro du diviseur, l'estim\'{e}e $[ \theta ]_e (t)$ appartient au
halo de $\theta$. Cette largeur est, alors, appr\'{e}ciable. Si, par
contre, cette largeur est infinit\'{e}simale, $[ \theta ]_e (t)$
n'appartient pas n\'{e}cessairement au halo de $\theta$.
\end{proposition}
\begin{remarque}
L'exp\'{e}rience pratique \og prouve \fg ~ qu'une \og courte \fg ~
largeur suffit, en g\'{e}n\'{e}ral, pour obtenir une bonne estimation.
\end{remarque}
\begin{remarque}
Supposons que le bruit soit stationnaire et blanc. Sa variance \'{e}tant
proportionnelle \`{a} la distribution de Dirac \`{a} l'origine, on dit ({\it
cf.} \cite{proakis}) qu'il contient toutes les fr\'{e}quences et que des
filtres passe-bas, comme les int\'{e}grales it\'{e}r\'{e}es de (\ref{estim}),
laissent passer les basses fr\'{e}quences. Renvoyons \`{a} \cite{hal} pour
lever ces contradictions apparentes.
\end{remarque}
\begin{remarque}\label{asympt}
Reprenons (\ref{discr}) et supprimons en (\ref{centlim}) la
condition $t_F \leq 1$. Pour $\Delta t \to 0$ et $t_F = \bar{N}^2 =
\frac{1}{(\Delta t)^2}$, la somme (\ref{centlim}) ne tend plus
presque s\^urement vers $0$. Nos m\'{e}thodes sont donc, par essence,
non asymptotiques.
\end{remarque}
\begin{remarque}
On ne peut quantifier, d'apr\`{e}s la proposition \ref{fen} et la
remarque \ref{asympt}, les {\em performances} d'un estimateur de
type (\ref{estim}) dans le cadre habituel ({\it cf.}
\cite{ibra,proakis}), qui est probabiliste, asymptotique et
hilbertien (fonctions de carr\'{e}s sommables). Des crit\`{e}res plus
appropri\'{e}s seront propos\'{e}s.
\end{remarque}

\subsubsection{D\'{e}modulation et erreurs en rafales}
En {\em d\'{e}modulation} ({\it cf.} \cite{code,proakis}), le param\`{e}tre
$\theta$ \`{a} estimer, appel\'{e} {\em symbole}, est associ\'{e}, en g\'{e}n\'{e}ral, \`{a}
un signal, comme un sinus, cardinal ou non, solution d'une \'{e}quation
diff\'{e}rentielle lin\'{e}aire, \`{a} coefficients polyn\^omiaux. Les erreurs
en rafales, tr\`{e}s courantes, que les codes correcteurs d'erreurs
doivent contrecarrer ({\it cf.} \cite{code,proakis}), s'interpr\`{e}tent
ainsi: l'estimation repose sur une hypoth\`{e}se fausse, \`{a} savoir le
caract\`{e}re centr\'{e} du bruit. Comme la dur\'{e}e de transmission d'un
symbole est \og courte \fg, on suppose que la moyenne $m(t)$, {\em a
priori} inconnue, est approch\'{e}e par un polyn\^ome $p(t)$ de degr\'{e}
limit\'{e}, donn\'{e}: $p(t)$ est l'ombre de $m(t)$. Il semblerait que l'on
puisse souvent supposer cette moyenne \og \`{a} peu pr\`{e}s \fg ~ constante
pendant chaque dur\'{e}e de transmission, ce qui simplifie les calculs.

\'Ecrivons ${\mathfrak{n}} = {\mathfrak{n}}_0 + p$ et consid\'{e}rons
$p$ comme une perturbation structur\'{e}e que l'on annihile
\cite{cras,fmmsr} avec une puissance limit\'{e}e, suffisante, de
$\frac{d}{dt}$. On obtient une formule analogue \`{a} (\ref{estim}) en y
rempla\c{c}ant ${\mathfrak{n}}$ par ${\mathfrak{n}}_0$. La possible
valeur du r\'{e}sultat suivant, de d\'{e}monstration semblable \`{a} celle de la
proposition \ref{fen}, tient au fait que l'estimateur doit d\'{e}tecter
un symbole parmi plusieurs, \og \'{e}loign\'{e}s \fg ~ les uns des autres.
\begin{proposition}
Supposons que l'ombre du signal ne soit pas annihil\'{e}e par une
puissance limit\'{e}e de $\frac{d}{dt}$. Il existe, alors, une estim\'{e}e
$[ \theta ]_e (t)$ du symbole appartenant au halo de $\theta$, pour
toute largeur $t$ n'appartenant pas au halo d'un z\'{e}ro du diviseur.
\end{proposition}

\vspace{0.2cm} \noindent{\small{\bf Remerciements}. L'auteur exprime
sa reconnaissace \`a C. Lobry (Nice) et T. Sari (Mulhouse) pour
d'utiles conversations.}

% Les remerciements sont dans une section, sans num\`{E}rotation
\selectlanguage{english}

\end{document}